\documentclass[fleqn,usenatbib]{mnras}

\usepackage{newtxtext,newtxmath}

\usepackage[T1]{fontenc}

\DeclareRobustCommand{\VAN}[3]{#2}
\let\VANthebibliography\thebibliography
\def\thebibliography{\DeclareRobustCommand{\VAN}[3]{##3}\VANthebibliography}
\DeclareRobustCommand{\DA}[3]{#2}
\let\DAthebibliography\thebibliography
\def\thebibliography{\DeclareRobustCommand{\DA}[3]{##3}\DAthebibliography}

\newcommand{\magphys}{\textsc{Magphys}\xspace}
\newcommand{\pros}{\textsc{Prospector}\xspace}

\usepackage{graphicx}	
\usepackage{amsmath}	
\usepackage{caption}
\usepackage{subcaption}
\usepackage{soul}
\usepackage{comment}
\usepackage{xcolor}
\definecolor{sdasgreen}{RGB}{30,150,60}
\usepackage{ulem}
\usepackage{xspace}

\title[Beware the recent past]{Beware the recent past: a bias in spectral energy distribution modeling due to bursty star formation}

\author[P.~Haskell et al.]{
P.~Haskell$^{1}$\thanks{E-mail: ph18aai@herts.ac.uk},
S.~Das$^1$,
D.~J.~B.~Smith$^{1},$
R.~K.~Cochrane$^{2,3},$
C.~C.~Hayward$^{3}$
and D.~Angl\'es-Alc\'azar$^4{}^,{}^3$
\\
$^{1}$Centre for Astrophysics Research, University of Hertfordshire, College Lane, Hatfield AL10 9AB, UK
\\
$^{2}$Department of Astronomy, Columbia University, New York, NY 10027, USA
\\
$^{3}$Center for Computational Astrophysics, Flatiron Institute, 162 Fifth Avenue, New York, NY 10010, USA
\\
$^{4}$Department of Physics, University of Connecticut, 196 Auditorium Road, U-3046, Storrs, CT 06269-3046, USA 
}

\date{Accepted XXX. Received YYY; in original form ZZZ}

\pubyear{2023}

\begin{document}
\label{firstpage}
\pagerange{\pageref{firstpage}--\pageref{lastpage}}
\maketitle

\begin{abstract}
We investigate how the recovery of  galaxy star formation rates (SFRs) using energy-balance spectral energy distribution (SED) fitting codes depends on their recent star formation histories (SFHs).
We use the \magphys\ and \pros\ codes to fit 6,706 synthetic spectral energy distributions of simulated massive galaxies at $1 < z < 8$ from the Feedback in Realistic Environments (FIRE) project.
We identify a previously-unknown systematic error in the \magphys\ results due to bursty star formation: the derived SFRs can differ from the truth by as much as 1 dex, at large statistical significance ($>5\sigma$), depending on the details of their recent SFH. SFRs inferred using \pros\ with non-parametric SFHs do not exhibit this trend. We show that using parametric SFHs (pSFHs) causes SFR uncertainties to be underestimated by a factor of up to $5\times$. Although this undoubtedly contributes to the significance of the systematic, it cannot explain the largest biases in the SFRs of the starbursting galaxies, which could be caused by details of the stochastic prior sampling or the burst implementation in the \magphys libraries. 
We advise against using pSFHs and urge careful consideration of starbursts when SED modelling galaxies where the SFR may have changed significantly over the last $\sim$100 Myr, such as recently quenched galaxies, or those experiencing a burst. This concern is especially relevant, e.g. when fitting \textit{JWST} observations of very high-redshift galaxies.

\end{abstract}




\section{Introduction}
\label{sec:Introduction}
Star formation rates (SFRs) and star formation histories (SFHs) are both of great importance to our understanding of galaxies. SFRs are critical since (along with their stellar mass) they enable us to put galaxies in their contemporary context relative to the evolving SFR--stellar mass relation \citep[e.g.][]{Noeske2007}. SFHs encode the build-up of stellar mass in the Universe \citep{Dye2008}, potentially yielding insight into the physical processes that drive galaxy evolution \citep[e.g.][]{Madau1996,Behroozi2013,Wang2022}. SFRs and other galaxy properties can be inferred for high-redshift galaxies by fitting a model spectral energy distribution (SED) to available data, which can include spectroscopy as well as photometry. There are many freely-available SED fitting codes able to do this \citep[as discussed by e.g.][]{Walcher2011,Conroy2013,Baes2019}.

SED models frequently differ in their approach and assumptions in important ways which potentially impact the fidelity of inferred galaxy properties (\citealt{Hunt2019}, \citealt{Pacifici2023}).
For example, they may employ different stellar population synthesis models (e.g. \citealt{Bruzual2003}, \citealt{Maraston2005},
\citealt{Conroy2009}; see also \citealt{Baldwin2018}), initial mass functions (IMFs; e.g. \citealt{Salpeter1955}, \citealt{Kroupa2001}, \citealt{Chabrier2003}; see also \citealt{Kroupa2019}),  dust models (e.g. \citealt{Calzetti1997},  \citealt{Charlot2000}, \citealt{Draine2007b}, \citealt{daCunha2008}) and SFH forms. These can be parametric or non-parametric; while the former typically use one of many simple functional forms to represent the SFH \citep[e.g.][]{Behroozi2013,Simha2014,Carnall2019}, the latter can be implemented e.g. by partitioning the history into time bins and allocating stellar mass to each bin from a chosen prior distribution as the observations allow (e.g. \citealt{Leja2019}). 

Many SED fitters, including those used in this study, assume balance between the energy absorbed by dust at short wavelengths and that re-radiated in the far-infrared. While this must be true overall,
strict equality should not hold for every line of sight since dust attenuation is non-isotropic. Recent observations of spatial offsets between the starlight and dust in high-redshift galaxies  \citep[e.g.][]{Hodge2016,Hodge2019,Cochrane2021} cast doubt on whether these codes are appropriate in such cases. However \citet[hereafter H23]{Haskell2023} showed that the performance of \magphys\ is similar for galaxies with UV\slash FIR spatial offsets as it is for local galaxies \citep{Hayward2015}; energy-balance codes remain the gold standard for deriving the properties of galaxies from photometric surveys. 

With myriad codes available, with different physics and underlying assumptions, it is critical to validate the performance of each method. As well as the works of this group, which have compared the physical properties inferred using energy-balance SED fitting with \magphys\ to simulated galaxies for which the `ground truth' properties are known (\citealt{Hayward2015,Smith2015,Smith2018}; H23), 
there is a large body of work by other authors aiming to understand how well SED fitting works in different ways (e.g. \citealt{daCunha2008}, \citealt{Lee2009}, \citealt{Noll2009}, \citealt{Wuyts2009}, \citealt{Lee2010}, \citealt{Pforr2012}, \citealt{Simha2014}, \citealt{Mobasher2015}, \citealt{Hunt2019}, \citealt{Dudzeviciute2020}, \citealt{Lower2020}, \citealt{Pacifici2023}, \citealt{Best2023}). Since validation tests are often accomplished by averaging over a large population of galaxies, there is always danger of ``washing out'' important details or systematic effects that could impact our understanding of how well SED fitting methods work in real-Universe situations. For example, recent papers have shown that non-parametric models are better able to recover model SFHs with recent sharp changes than fitters using parametric SFHs (hereafter pSFHs; e.g. \citealt{Leja2019}, \citealt{Carnall2019}, \citealt{Seuss2022}, \citealt{Narayanan2023}).

Here, we build on H23 and run the \pros SED fitting code on the same synthetic galaxy SEDs for which we already have \magphys\ fits, focusing on how the inferred SFR depends on the recent SFH within the last 100 Myr. Galaxies experiencing rapid changes in SFR over short ($\sim$10s-100s of Myr) timescales have been identified both observationally (e.g. \citealt{Guo2016}, \citealt{Faisst2019},  \citealt{Broussard2022}, \citealt{Looser2023}, \citealt{Wang2023}, \citealt{Woodrum2023}) and in simulations (e.g. \citealt{Sparre2017}, \citealt{Iyer2020}, \citealt{FloresVelasquez2021}, \citealt{Hopkins2023}, \citealt{Narayanan2023}, \citealt{Dome2024}). In the FIRE simulations, massive galaxies at high redshift and dwarfs at all redshifts exhibit very bursty SFHs, with short ($\sim$10-Myr) bursts followed by temporarily quenched periods\footnote{We will use the shorthand `temp-quenched' to refer to the periods of low-specific SFR between bursts that are common in the FIRE simulations analysed here. 
We use the term `starburst' to refer to simulated galaxies that are currently undergoing a burst, i.e.\,the 10 Myr-averaged SFR is greater than the 100 Myr-averaged value.} lasting a few 100 Myr; see \citet{Sparre2017} for a detailed analysis and \citet{Hayward2017} and \citet{Hopkins2023} for discussions of the physical origin of this behavior. They thus provide an excellent test data set for exploring the effects of bursty star formation on SED modeling inferences.\footnote{Owing to the lack of AGN feedback in these simulations, none of the simulated galaxies are permanently quenched. However, since the bias depends on the SFH over the past 100 Myr, our conclusions should apply equally to permanently- as well as temp-quenched galaxies.} 

In this work, we demonstrate that \magphys\ results exhibit a systematic bias in the inferred SFR: the SFR is overestimated (underestimated) for simulated galaxies experiencing a burst (temporarily quenched period). We also show that \pros SFRs do not exhibit this bias.
This Letter is structured as follows. In Section \ref{sec:Method} we outline the tools and methods used to create the observations and to fit the SEDs, while in Section \ref{sec:Results} we present the results of the fitting, and in Section \ref{sec:Discussion} discuss the implications. In Section \ref{sec:conclusions} we make some concluding remarks. We adopt a standard cosmology with $H_0=70\,$km\,s$^{-1}$\,Mpc$^{-1}$, $\Omega_M=0.3$, and $\Omega_\Lambda=0.7$.

\begin{figure}
\centering
\includegraphics[width=0.45\textwidth]{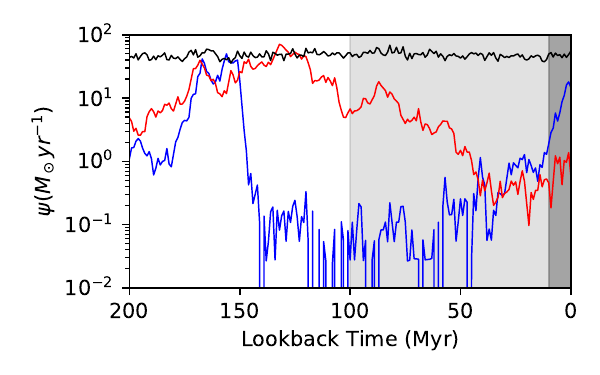}
\caption{The recent SFHs (past 200 Myr) of three example simulated galaxies, showing a range of $\eta$ values. The dark grey band encloses the previous $10\,\mathrm{Myr}$, while the lighter band encloses the past $100\;\mathrm{Myr}$; $\eta$ is the ratio of the SFRs averaged over those two time periods. The black line shows an example of constant star formation with $\eta \approx 0$, while the red line shows a `temporarily quenched' case ($\eta=-0.64$). The blue line shows an example simulated galaxy observed during a burst, where the star formation activity averaged over the previous $10\;\mathrm{Myr}$ has increased compared to the average over the previous $100\,\mathrm{Myr}$ (giving $\eta=0.88$).} 
\label{fig:SFR_examples}
\end{figure}

\section{Method}
\label{sec:Method}

To provide the synthetic SEDs used for this test, we used four cosmological zoom-in simulations of massive high-redshift galaxies from the FIRE project.\footnote{\url{http://fire.northwestern.edu}} The halos were first presented in \citet{Feldmann2016}, and subsequently resimulated by \citet{Angles2017} using the FIRE-2 code \citep{Hopkins2018}; we use the simulations from that work. The simulations use the code \textsc{gizmo} \citep{Hopkins2015},\footnote{\url{http://www.tapir.caltech.edu/~phopkins/Site/GIZMO.html}} with hydrodynamics solved using the mesh-free Lagrangian Godunov `MFM' method. The simulations include cooling and heating from a meta-galactic background and local stellar sources from $T\sim10-10^{10}\,$K; star formation in locally self-gravitating, dense, self-shielding molecular, Jeans-unstable gas; and stellar feedback from OB \&\ AGB mass-loss, SNe Ia \&\ II, and multi-wavelength photo-heating and radiation pressure, with inputs taken directly from stellar evolution models. The FIRE physics, source code, and all numerical parameters are 
identical to those in \citet{Hopkins2018}.

Snapshots of the halos' central galaxies were taken from $1<z<8$ at intervals of $15 - 25\,\mathrm{Myr}$. For each snapshot, we compute the true $10\,\mathrm{Myr}$ and $100$ Myr-averaged SFRs, $\overline{\psi}_\mathrm{10\,Myr}$ and $\overline{\psi}_\mathrm{100\,Myr}$, respectively. Following \citet{Broussard2019}, we quantify the recent star forming activity using the `burst indicator'  $\eta = \log_{10}(\mathrm{\overline{\psi}_{10\,\mathrm{Myr}}}/\mathrm{\overline{\psi}_{100\,\mathrm{Myr}}})$, such that 
\[
    \eta 
\begin{cases}
    <0,& \text{galaxy has reduced SFR within the last $10\,\mathrm{Myr}$}; \\
    \approx 0,& \text{galaxy has had a constant SFR over the last $100\,\mathrm{Myr}$};\\
    >0,& \text{galaxy has increased SFR in the last $10\,\mathrm{Myr}$}.\\
\end{cases}
\]
Figure \ref{fig:SFR_examples} shows three example SFHs of snapshots used in this study with $\eta\approx0$ (black line), $\eta<0$ (red line) and $\eta>0$ (blue line). 

\citet{Cochrane2019} calculated synthetic SEDs for these simulated galaxies using the radiative transfer code SKIRT.\footnote{\url{http://www.skirt.ugent.be/}}
Each snapshot was `observed' at 7 viewing angles spaced at 30$^{\circ}$ intervals, ranging from aligned with the angular momentum vector to anti-aligned. This resulted in 6,706 forward-modeled SEDs. See \citet{Cochrane2019,Cochrane2022,Cochrane2023a,Cochrane2023b} and \citet{Parsotan2021} for further details about the SKIRT calculations and other applications of these and similar simulations/SEDs.
We convolved these SEDs with the filter response curves for 18 photometric bands spanning observed-frame wavelengths $0.39 < \lambda \slash \mu \mathrm{m} < 500$ (we refer the interested reader to H23 for further details) 
assuming a signal-to-noise ratio of 5 in every band \citep[following][]{Smith2018}.
We fit the synthetic SEDs using two SED fitting codes, \magphys\ \citep[hereafter DC08]{daCunha2008} and \pros (\citealt{Leja2017,Leja2019}, \citealt{Johnson2021}), fixing the redshift to the true value (i.e.\,photometric redshift errors do not contribute to the bias demonstrated here). We use the high-redshift version of \magphys\ \citep{daCunha2015} which builds model UV--FIR SEDs by linking a stellar library containing 50,000 pre-computed SFHs with those which satisfy the energy balance criterion among another pre-computed library of 25,000 \citet{Charlot2000} dust models; see DC08 and H23 for detailed discussions. By calculating the $\chi^2$ goodness-of-fit parameter between the observed photometry and every combination which satisfies the energy balance, \magphys\ is able to marginalise over the prior distribution to estimate posterior probability distributions for each property in the model. \magphys\ assumes an initial mass function (IMF) from \citet{Chabrier2003}, and stellar models from \citet{Bruzual2003} for metallicity between 0.02 and 2 times solar.

For the \pros fits, we followed the process described in Das et al., {\it in preparation}. \pros uses the Flexible Stellar Populations Synthesis \citep[FSPS;][]{Conroy2009} code assuming a \cite{Kroupa2001} IMF and solar metallicity. Dust attenuation is modelled using the two-component \cite{Charlot2000} model and the \cite{Kriek2013} attenuation curve. The three-parameter \citet[hereafter DL07]{Draine2007b} dust emission templates are used to model the shape of the IR SED. The dynamic nested sampling library \texttt{Dynesty} \citep{Speagle2020} is used to estimate Bayesian posteriors and evidences.  

For both fitters, we consider an SED fit to be acceptable if the best fit $\chi^2$ is below the 99 per cent confidence limit taken from standard $\chi^2$ tables with the degrees of freedom calculated following \citet[who calculated this only for \magphys; here we assume that it applies equally well to the \pros results, though this is perhaps unlikely to be true given the differences between the two sets of models]{Smith2012b}. Throughout the subsequent analysis, we consider only those SEDs with satisfactory fits, though our conclusions are qualitatively unchanged if we also consider the ``bad fits''. 

Based on these analyses, we obtained five sets of SFR estimates for each snapshot. The \magphys\ SFRs assume a fixed parameterisation of the SFH based on a parametric model overlaid with random starbursts (as detailed in DC08). In addition, we have four sets of results from \pros, assuming different SFH priors: parametric, Dirichlet, continuity, and bursty continuity. The parametric prior assumes a delayed  exponentially declining SFH, and does not account for bursts.\footnote{Using \pros with bursts is not recommended in the documentation.}  The Dirichlet prior assumes that the fractional specific star formation rate (sSFR) in each time bin follows a Dirichlet distribution, with a concentration parameter, $\alpha$, that controls the shape of the SFH \citep{Leja2017,Leja2019}, and which we set to unity. The continuity prior directly fits for $\Delta\log(\mathrm{SFR})$ between neighbouring time bins using a Student's $t$-distribution with scale factor $\sigma = 0.3$ and two degrees of freedom. This prior discourages abrupt changes in SFR between adjacent bins but remains flexible enough to fit both star-forming and quenched galaxies \citep{Leja2019,Johnson2021}. Following \cite{Tacchella2022B}, we also consider the bursty continuity prior -- a modified form of the continuity prior with $\sigma=1$ and $\nu=2$, which allows greater variability in the SFH. For each of these priors, we adopt the following binning scheme: the first two and the last bins are kept the same for all sources, covering $0 < t_l < 10$ Myrs, $10 < t_l < 100$ Myrs, and $0.85t_z < t_l < t_z$, respectively. Here, $t_z$ represents the age of the Universe at the object's redshift, and $t_l$ is the lookback time. The remaining time between 100 Myrs and $0.85t_z$ is evenly spaced in logarithmic time. The number of bins is chosen to ensure that $\log_{10}(\mathrm{bin\: width\slash\mathrm{Gyr}}) > 0.02$, ranging from eight bins (for the lowest redshifts in our simulations), to five (for the highest).

To quantify the fidelity of the $100\,\mathrm{Myr}$ averaged SFR inferred using \magphys\ and \pros, following H23, we use the residual between the inferred SFR and the true value from the simulation: $\Delta \log_{10} \psi = \log_{10} (\psi_\mathrm{inferred}) - \log_{10} (\psi_\mathrm{true})$.

\section {Results}
\label{sec:Results}

Table \ref{tab:residuals} shows the fit success rate and the average SFR residual for each of the five runs. It is immediately apparent that \pros is able to produce acceptable fits for a larger fraction of the snapshots ($>90$ per cent fit success rate irrespective of the prior assumed) than \magphys\ (83\,per cent). This may reflect the effect that the \magphys\ pre-computed libraries -- coupled with the requirement that it consider only SFHs shorter than the age of the Universe at a given redshift -- result in the priors becoming increasingly poorly-sampled at higher redshifts, and acceptable \magphys\ fits increasingly hard to obtain at earlier epochs. 
Secondly, we note that the typical SFR uncertainties for the non-parametric runs shown in Table \ref{tab:residuals} are significantly larger than those yielded by the pSFH runs (with both \magphys and \pros) and we will return to this point below. These details notwithstanding, it is clear from Table \ref{tab:residuals} that there is no significant difference in the overall $\Delta \log_{10} \psi$ once the uncertainties are considered, meaning that both codes do a similarly acceptable job of recovering the true SFR on average. 

\begin{table}
 \centering
 \caption{SED fit success rate and typical SFR estimate fidelity, $\Delta \log_{10}(\psi)$, for the \magphys\ and \pros fits to our synthetic observations. For the four sets of \pros fits, the different SFH priors are indicated.} 
 \label{tab:filter_details }
\begin{tabular}{lcl}
\cline {1-3}
Fitter/prior&\% Fit success&$\Delta\log_{10}(\psi)$\\
\cline {1-3}
\magphys (pre-defined) &83& -0.11 $\pm$ 0.06\\
\pros (pSFH; delayed $\tau$ model) &92&-0.11 $\pm$ 0.05\\
\pros (Bursty-continuity) &91&-0.09 $\pm$ 0.24\\
\pros (Dirichlet, $\alpha=1$) &92&-0.30 $\pm$ 0.19\\
\pros (Continuity) &92&-0.12 $\pm$ 0.12\\
\cline {1-3}
\end{tabular}
\label{tab:residuals}
\end{table}

\begin{figure*}
\centering 
\includegraphics[width=\textwidth]{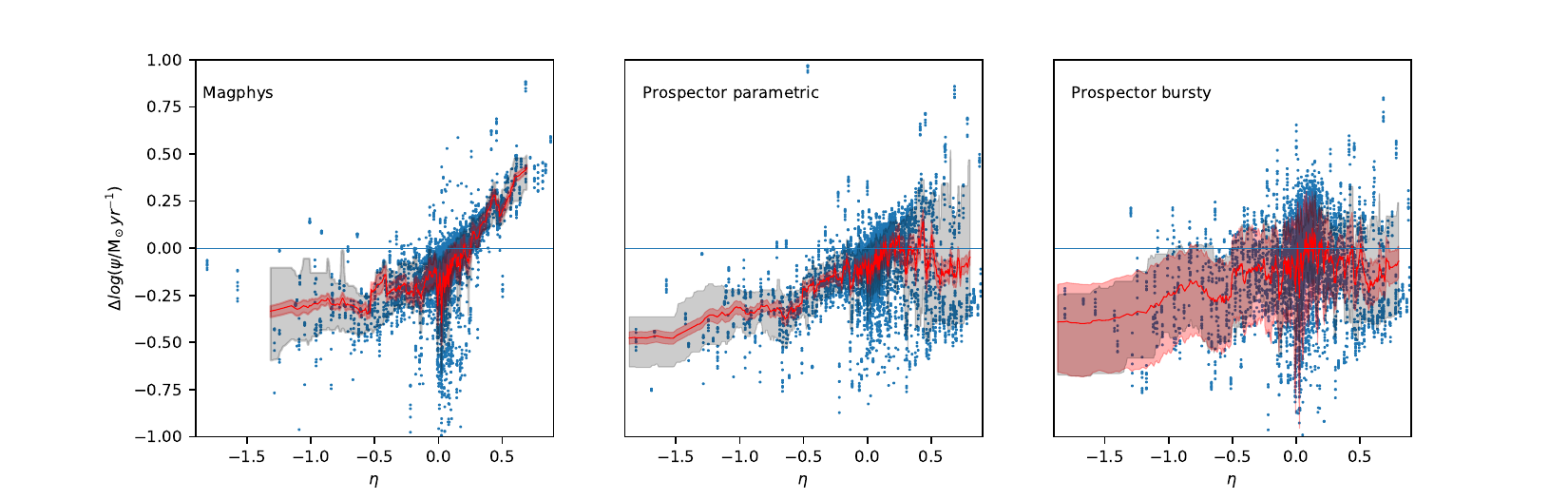}
\caption{The LH panel shows the difference between the inferred and true SFRs averaged over the last 100\,Myr, $\Delta \log_{10} \psi = \log_{10} (\psi_\mathrm{inferred}) - \log_{10} (\psi_\mathrm{true})$, as a function of the burst indicator $\eta \equiv \log_{10}(\mathrm{\overline{\psi}_{10\,\mathrm{Myr}}}/\mathrm{\overline{\psi}_{100\,\mathrm{Myr}}})$. The centre and RH panels show the same obtained using \pros with the pSFH and bursty continuity SFH prior. In all panels the red line shows the mean residual averaged over 100 data points with the red shaded area enclosing the averaged 16th and 84th percentile residuals from the PDF. The grey shaded area indicates the 16th and 84th percentiles of the scatter in $\Delta \log_{10} \psi$ within each averaged point, and the blue markers show values for individual SEDs. The \magphys\ SFRs exhibit a systematic bias: the SFR tends to be underestimated for `temp-quenched' galaxies and overestimated for galaxies that have experienced a burst within the past 10\,Myr.}
\label{fig:SFR_against_SFH}
\end{figure*}

We now investigate the fidelity of the \magphys\ and \pros SFR estimates as a function of the recent SFH. In Figure \ref{fig:SFR_against_SFH}, we plot $\Delta\log(\psi)$ against $\eta$, with \magphys\ results shown in the  left-hand panel, \pros pSFH results in the centre panel, and the \pros  bursty continuity prior results in the right-hand panel. The axes are chosen such that `starburst' snapshots appear to the right of each plot, and `temp-quenched' ones to the left. The values for individual SEDs are shown as points, with red lines showing a running average of $\Delta \log_{10}(\psi)$ for 100 data points. The red areas enclose the 16th and 84th percentiles of the inferred SFR PDF using the same running average, while the grey shading encloses the 16th and 84th percentiles of the scatter within each corresponding subsample. 

Two effects are immediately apparent. Firstly, there is a systematic trend in the \magphys\ SFRs, such that while they are typically overestimated by $\sim 0.4$\,dex at $\eta > 0.5$ (the `starburst' snapshots), the converse is true at $\eta < -0.5$ (the `temp-quenched' snapshots) where \magphys\ typically underestimates the true SFRs by $\sim0.3$\,dex. This is clearly of concern, since we know that both `starburst' and `temp-quenched' galaxies exist in the real Universe \citep[e.g.][]{Guo2016,Faisst2019,Broussard2022,Looser2023} and this effect could result in a systematic error in the SFRs of galaxies that could be 1 dex in magnitude in the worst cases. Secondly, and underscoring the severity of the first issue, there is a clear tendency for the typical pSFH uncertainties to underestimate the degree of scatter visible in the data points (grey shading). This effect becomes increasingly noticeable as $|\eta|$ increases. 

The equivalent \pros\ bursty continuity plot shows the SFR recovery is approximately independent of $\eta$ with reasonable uncertainties at $\eta_\mathrm{true} < 0.5$ (where $\sim 90$ per cent of the snapshots lie). At these $\eta$, the $\sim 0.3$\,dex offset in the \pros SFRs relative to the ground truth is similar to that returned by \magphys\ at $\eta < -0.5$; however, in the \pros non-parametric results the deviation from zero is not statistically significant due to the realistic error bars (and we obtain similar results with the other non-parametric SFHs).

\section{Discussion}
\label{sec:Discussion}

Fundamental to SED fitting codes' abilities to recover the true properties of galaxies are (i) wavelength sampling (including range covered, and resolution),  (ii) priors, and (iii) sensitivity of the available observations. The first of these is closely related to the extent of our ability to estimate $\eta$ using SED fitting \citep[since different wavelengths have different ``response functions'' between the luminosity and the SFH; e.g.][]{kennicutt2012,Sparre2017}. In this work we focus solely on our ability to recover 100\,Myr-averaged SFRs using an example 18 bands of photometry similar to that used in \citet{Smith2021} and \citet{Best2023} to fit galaxies in the LOFAR deep fields \citep{Tasse2021,Sabater2021,Kondapally2021,Duncan2021}. This good wavelength sampling, in addition to the fact that we have not added noise to photometry, means that this test represents a best-case scenario: the bias that we have identified is fundamental, not due to noise or poor SED sampling, and thus cannot be addressed with `better data'.

\begin{figure}
\centering
\includegraphics[width=0.85\columnwidth]{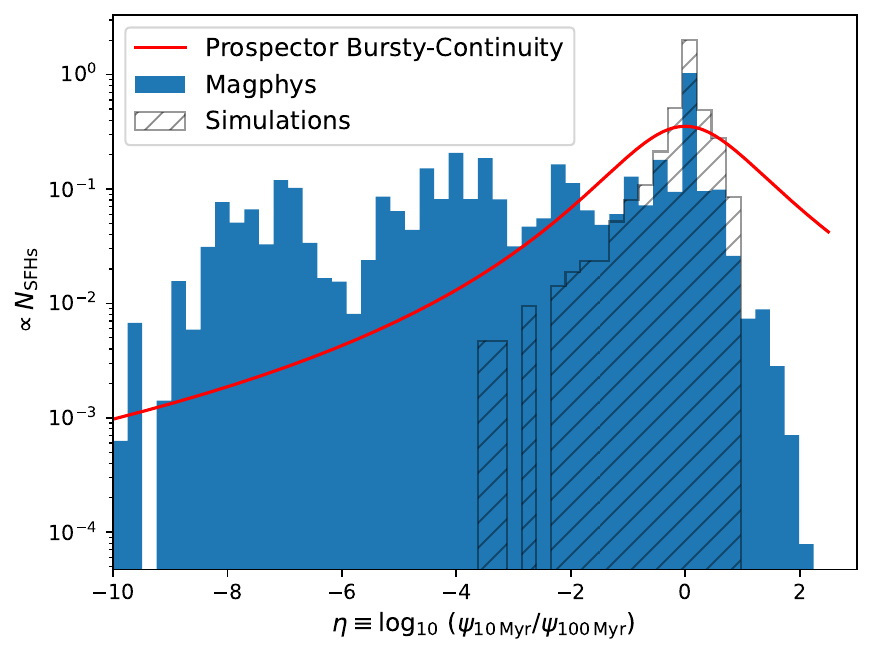}
    \caption{The distributions of $\eta$ values for 50,000 SFHs in the \magphys\ high-$z$ libraries (in blue), overlaid with the ground truth values from the simulations (hatched) and for the \pros bursty continuity prior in red}. Note the significantly larger $\eta$ range relative to Figure \ref{fig:SFR_against_SFH}.
    \label{fig:eta_magphys}
\end{figure}

In the \magphys model, the priors are immutable since it uses pre-computed SFH libraries with fixed sampling. The high-$z$ \magphys\ SFHs assume a delayed exponential parametric form, with random bursts superposed in such a manner that during the previous 2\,Gyr, 75 per cent of SFHs include a burst lasting between 30-300\,Myr and forming between 0.1 and 100 times the stellar mass formed by the underlying distribution \citep{daCunha2015}. 
The $\eta$ distribution for the \magphys\ high-$z$ SFHs is shown in Figure \ref{fig:eta_magphys}, overlaid with the distribution of values for the true SFHs of the simulated galaxies and for the assumed \pros bursty continuity prior. It is clear that both the \magphys\ libraries and the \pros bursty continuity prior allow for a significantly broader range of $\eta$ values than the simulations contain, so the prior on $\eta$ is not the source of the bias.

In Section \ref{sec:Results} we noted that the pSFH SFR uncertainties are increasingly underestimated as $|\eta|$ moves away from 0. 
Alongside the similarity in the mean $\Delta \psi$ between the \pros pSFH and non-parametric plots, this leads us to speculate that while the choice of prior may affect the uncertainties and crucially the statistical significance of the systematic trend, the cause of the largest upturn in the \magphys $\Delta \log_{10}\psi$ for the most bursty galaxies lies elsewhere.

\section{Conclusions}
\label{sec:conclusions}

We have used four cosmological zoom-in simulations spanning $1<z<8$, together with the radiative transfer code SKIRT, to generate 6,706 synthetic galaxy SEDs. We attempted to recover their true SFRs based on fitting the synthetic observations using \magphys\ and \pros. Although there is little difference between the fidelity of the recovered $\overline{\psi}_{100\,\mathrm{Myr}}$ values averaged across all of the simulated recovery tests using the two codes, we find that the accuracy of the \magphys-inferred SFRs is strongly dependent on the recent SFH. This trend is sufficiently large that the \magphys\ 100\,Myr-averaged SFRs of individual galaxies could be overestimated (underestimated) by as much as 1\,dex in the worst cases for galaxies experiencing starbursts (temporarily quenched periods). The trend is not evident in the 100\,Myr SFRs from \pros, and is significantly weakened in \pros results assuming parametric SFHs (pSFHs). We therefore speculate that this effect could be due to \magphys's burst recipe or the use of pre-defined libraries with fixed parameter sampling which may be unsuitable for some galaxies. Given this bias, and the significantly underestimated SFR uncertainties produced using pSFHs, we urge caution when employing pSFHs and\slash or pre-defined libraries to study the SEDs of observed galaxies that are likely to have bursty SFHs, such as high-redshift galaxies observed with \textit{JWST}.

\section*{Acknowledgements}

PH and DJBS thank the Flatiron Institute for their hospitality. DJBS and SD acknowledge support from UK STFC grants ST/V000624/1 and ST/W507490/1 respectively. This research has made use of NASA's Astrophysics Data System Bibliographic Services.

\section*{Data Availability}
\noindent Data will be shared on reasonable request to the first author.


\bibliographystyle{mnras}
\bibliography{library} 


\bsp	
\label{lastpage}
\end{document}